\documentclass[prl,aps,twocolumn,twoside,superscriptaddress]{revtex4}

\usepackage{amssymb}
\usepackage{amsmath}
\usepackage{amscd}
\usepackage{latexsym}
\usepackage{epsfig}
\usepackage{graphicx}
\usepackage{bbm}

\font\gensymbols=drgen10
\def\male{{\gensymbols\char"1A}}
\def\female{{\gensymbols\char"19}}

\newtheorem{defi}{Definition}

\newtheorem{exempel}[defi]{Example}

       \def\CE{{\cal E}}

       \def\CN{{\cal N}}

\def\beq{\begin{equation}}
\def\eeq{\end{equation}}

\newcommand{\tr}{{\operatorname{Tr}}}

\newcommand{\ket}[1]{{|{#1}\rangle}}

\newcommand{\1}{{\openone}}

\def\proj#1{ | #1 \rangle \langle #1 |}

\newlength{\blank}
\newlength{\equalsign}
\begin{document}

\title{A family of quantum protocols}
\author{Igor Devetak}
\email{devetak@us.ibm.com}
\affiliation{IBM T.~J.~Watson Research Center, PO Box 218, Yorktown Heights, NY 10598, USA}
\author{Aram W. Harrow}
\email{aram@mit.edu}
\affiliation{MIT Physics Dept., 77 Massachusetts Avenue, Cambridge, MA 02139, USA}
\author{Andreas Winter}
\email{a.j.winter@bris.ac.uk}
\affiliation{Department of Mathematics, University of Bristol,
     University Walk, Bristol BS8 1TW, UK}

\begin{abstract}
We introduce three new quantum protocols involving noisy quantum
channels and entangled states, and relate them operationally and
conceptually with four well-known old protocols.  Two of the new
protocols (the ``mother'' and ``father'') can generate the other five
``child'' protocols by direct application of teleportation and
super-dense coding, and can be derived in turn by making the old
protocols ``coherent.''  This gives very simple proofs for two famous
old protocols (the hashing inequality and 
quantum channel capacity) and provides the basis for optimal
tradeoff curves in several quantum information processing tasks.
\end{abstract}

\maketitle

\paragraph{Introduction.}
The central task of quantum information theory is to determine the
rates at which the quantum state of any physical object can be
transmitted from one location to another.  So far quantum information
theory incorporates a number of basic coding theorems, including
quantum compression \cite{schu}, and expressions for classical
\cite{HSW} and quantum \cite{Lloyd:Q, shor:Q, devetak} capacities of
quantum channels. In \cite{CR}, these results were formulated in terms
of asymptotic inter-conversion between information processing
resources, such as uses of a quantum channel, shared entanglement and
so on.  For instance, channel coding may be viewed as converting a
noisy channel into a noiseless one on a smaller input space.  A
particularly important class of problems in quantum information theory
involves converting a noisy quantum channel or shared noisy
entanglement between two spatially separated parties (conventionally
denoted by Alice and Bob), into a noiseless one, via local operations
possibly assisted by limited use of an auxiliary noiseless resource
such as a perfect qubit channel, shared ebits or one-way classical
communication.  Previously, this class of problems had only been
addressed as a collection of special cases, each requiring its own
complicated proof techniques to address.
In this Letter we consider basic protocols for each
member of this class, three of which are new, and observe that they
are naturally organized into two mutually dual hierarchies. This
result significantly simplifies the quantum information processing
landscape, revealing connections between scenarios previously thought
independent.  Some of our connections give constructive methods for
turning one protocol into another, so that a coding scheme for one
protocol yields codes for a whole class of other protocols.
Moreover, these basic protocols will provide the crucial
ingredient for constructing \emph{optimal} protocols and two
dimensional trade-offs.

\paragraph{The family of resource inequalities.}
\label{sec:family}
The following notation for information processing resources was proposed in \cite{CR}.
A noiseless qubit channel, noiseless classical bit channel 
and pure ebit (EPR pair) were denoted by $[q \rightarrow q]$, 
$[c \rightarrow c]$ and $[q \, q]$, respectively, 
reflecting their classical/quantum and dynamic/static nature.
A noisy bipartite state $\rho^{AB}$ is denoted by $\{q \, q\}$,
and a  general quantum channel 
$\CN: {\cal H}_{A'} \rightarrow {\cal H}_B$ is denoted by $\{q \rightarrow q\}$.
In either case one may define a class of pure states $\ket{\psi}^{ABE}$.
In the former, it consists of the purifications of $\rho^{AB}$, i.e.,
$\rho^{AB} = \tr_{E} \,{\psi}^{ABE}$.
In the latter, it corresponds to the outcome of sending half of
some $\ket{\phi}^{AA'}$ through the channel's Stinespring \cite{stinespring}
extension  
$U_\CN:{\cal H}_{A'} \rightarrow {\cal H}_B\otimes{\cal H}_E$ 
($\CN$, mapping states on $A'$ to states on $B$,
is obtained as the isometry $U_\CN$ followed by the partial trace over $E$.)
One may define the usual entropic quantities with respect to
the state  $\ket{\psi}^{ABE}$. Recall the definition
of the von Neumann entropy 
$H(A) = H(\psi^A) = -\tr (\psi^A \log \psi^A)$, 
where $\psi^A = \tr_{BE} \,\ket{\psi}^{ABE}$.
Further define the quantum mutual information \cite{adami:cerf}
$I(A;B)=H(A)+H(B)-H(AB)$
and the coherent information \cite{coherent}
$I_c(A\,\rangle B)=-H(A|B)=H(B)-H(AB)$;
the latter notation is from \cite{devetak:winter}.
Relative to the \emph{pure} state $\ket{\psi}^{ABE}$, $H(AB)=H(E)$ and
$H(AE)=H(B)$, so 
\begin{align*}
  \frac{1}{2} I(A;B) + \frac{1}{2} I(A;E) &= H(A), \\
  \frac{1}{2} I(A;B) - \frac{1}{2} I(A;E) &= I_c(A\,\rangle B).
\end{align*}
\par
It is possible to give meaning to {inequalities} between
the various resources with entropic quantities as coefficients. 
Consider, for instance, the 
 ``mother'' \emph{resource inequality} (RI):
\begin{equation}
  \frac{1}{2} I(A;E) \, [q \rightarrow q] + \{q \, q \}
                          \geq \frac{1}{2} I(A;B)\,[q \, q]. \tag{\female}
\end{equation}
It embodies an achievability statement: for any $\epsilon,\delta>0$,
for sufficiently large $n$ there exists
a protocol that uses up $n$ instances of a noisy bipartite state $\rho^{AB}$ 
and $\leq n \, (I(A;E)/2 + \delta)$ instances of a noiseless qubit channel, to
produce a state within trace distance $\epsilon$ of $\geq
n\,(I(A;B)-\delta)/2$ ebits.
The entropic quantities implicitly refer to any
$\ket{\psi}^{ABE}$ associated with the noisy resource $\rho^{AB}$.
The resources on the left (right) hand side are called input (output) resources,
respectively. We defer the construction of such a protocol to the next section.
%
\par
As we shall see, there exists a  dual ``father'' RI, related to the mother by
replacing dynamic resources with static ones and vice versa:
\begin{equation}
  \frac{1}{2} I(A;E) \, [q \, q] + \{q  \rightarrow q \}
                          \geq \frac{1}{2} I(A;B)\,  [q  \rightarrow q]. \tag{\male}
\end{equation}
Again, it means that for sufficiently large $n$ there exists
a protocol which uses $n$ copies of $\CN$ assisted by 
$\approx n \, I(A;E)/2$ ebits of entanglement to
simulate arbitrarily faithfully the effect of  
$\approx n\,I(A;B)/2$ noiseless qubit channels.
The entropic quantities implictly refer to any
$\ket{\psi}^{ABE}$ associated with the noisy resource $\CN$.

Note that in the noiseless case (pure ebit or perfect qubit channel)
both parents express trivial identities.
\par
We shall combine them with the activating noiseless resource
inequalities corresponding to 
teleportation \cite{tp}
\begin{equation}
  2\,[c \rightarrow c] + [q \, q] \succeq [q \rightarrow q] \tag{TP}
\end{equation}
and super-dense coding \cite{sd}
\begin{equation}
  [q \rightarrow q] + [q \, q] \succeq 2  \,[c \rightarrow c], \tag{SD}
\end{equation}
to generate their offspring.
Here we use ``$\succeq$'' to denote exact achievability
(as opposed to the asymptotic ``$\geq$'').
\par
They may be applied to a parent RI by either prepending 
(the output of TP/SD is used as an input to a protocol implementing 
the parent RI) or appending (the output of the parent is used as an input to TP/SD).
In addition to (TP) and (SD), we shall also make use of 
a third noiseless RI given by 
\begin{equation}
[q \rightarrow q] \succeq  [q \, q]  \tag{QE}.
\end{equation}
It is trivially implemented by sending half of an EPR pair through the
qubit channel. 
\par
\begin{figure}[ht]
  \includegraphics[width=7.5cm]{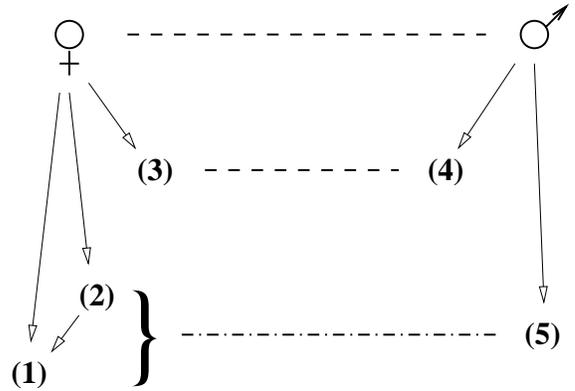}
  \caption{The family tree: the dashed lines signify duality,
    and the dashed-dotted line is the almost-duality described in the text.
    The solid arrows signify descendance via (TP), (SD) or (QE).}
  \label{fig:babushka}
\end{figure}

Each parent has her or his own children,
as shown in Fig.~\ref{fig:babushka}.
Let us consider the mother first; she has three children.
The first one is a new RI,
a noisy version of teleportation suggested to us by
Burkard \cite{guido}, in which noisy entanglement is combined with
classical communication to teleport a quantum state.  It is
obtained by appending (TP) to the mother:
\beq
  I(A;B) \, [c \rightarrow c] + \{q \, q \} \geq I_c(A\,\rangle B) \,[q \rightarrow q].
  \label{eq1}
\eeq
\par
The second is the recently proved ``hashing inequality'' \cite{devetak:winter}
(including the classical communication cost) which is known to yield
the optimal one-way distillable entanglement.
It follows from prepending (TP) to the mother:
\beq
  I(A;E) \, [c \rightarrow c] + \{q \, q \} \geq I_c(A\,\rangle B) \,[q \, q].
  \label{eq2}
\eeq
Note that (\ref{eq2}) also yields (\ref{eq1}), by appending (TP).
\par
The third is a noisy version of super-dense coding, which first
appeared (somewhat disguised) in \cite{H3LT} and is  
obtained by appending (SD) to the mother:
\beq
  H(A) \, [q \rightarrow q] + \{q \, q \} \geq I(A; B) \,[c \rightarrow c].
  \label{eq3}
\eeq
The father  doesn't quite make it to three children: he has only two.
Appending (SD) to him gives the coding for entanglement-assisted
classical information transmission \cite{BSST}:
\beq
  H(A) \, [q \, q] + \{q \rightarrow q \} \geq I(A; B) \,[c \rightarrow c].
  \label{eq4}
\eeq
Note that it is dual to (\ref{eq3}), at least as far as the quantum parts
are concerned. 
\par
There's one more thing we can do:
append (QE) to to a fraction of the output of 
(\male) to recover the famous quantum channel capacity result 
\cite{Lloyd:Q,shor:Q,devetak}
\beq
  \{q \rightarrow q\} \geq I_c(A\,\rangle B) \,[q  \rightarrow q].
  \label{eq5}
\eeq
This one is almost dual to (\ref{eq2}), and can be
made formally dual by wasting $I(A; E)\,[c \rightarrow c]$. 

The reason that the mother-father duality does not propagate
perfectly down the family tree lies in the lack of duality 
between (TP) and (QE). While (SD) is self-dual under the interchange of
$[qq]$ and $[q \rightarrow q]$, (TP) and (QE) become mutually dual only
by wastefully adding $2[c \rightarrow c]$ to the left hand side of (QE).
In this light, even (\ref{eq1})
has a dual RI: a rather wasteful version of (\ref{eq5}).

\paragraph{Coherent communication.}
\label{sec:ccc}
Having demonstrated the power of the parent resource inequalities,
we now address the question of constructing protocols implementing them.
Recently, the importance of 
``coherent communication'' was recognized \cite{aram}:
a \emph{coherent bit channel} is defined as the isometric mapping
\beq
  \ket{x}^A \mapsto \ket{x}^A \ket{x}^B
  \label{map}
\eeq
for a basis $\{ \ket{x}: x \in \{0,1\} \}$ of the qubit system $A$. 
Note that this transformation implements a noiseless transmission of 
the classical index $x$, but may also be used to create entanglement
by applying it to superpositions of $\ket{0}$ and $\ket{1}$.
Viewed as a resource we shall denote it by $[q \rightarrow q \, q] $.
In what follows it shall often be used in lieu of
the classical bit channel $[c \rightarrow c]$.
\par
In \cite{aram} it is shown that (SD) can be made ``coherent''
to yield two coherent bits 
$$[q \rightarrow q] + [q \, q] \succeq 2  \,[q \rightarrow q \, q] .$$
On the other hand, using coherent bits for teleportation has
the virtue of creating entanglement as a by-product
$$2\,[q \rightarrow q \, q]  + [q \, q] \succeq [q \rightarrow q] + 2\,[qq].$$
Hence we have the equivalence, modulo catalytic entanglement (symbolized by
the superscript $c$),
$$2 [q \rightarrow q \, q]  \stackrel{c}{\equiv} [q \rightarrow q] + [q \, q],$$
which gives us the asymptotic equivalence \cite{aram},
\begin{equation}
  \label{eq:ccc}
  [q \rightarrow q \, q]  = \frac{1}{2} \left( [q \rightarrow q] + [q \, q] \right).
\end{equation}
\par
Note that in the previous section we have already made use of the
fact that recycling allows us to convert catalytic formulas
(i.e., cancellation of equal terms left and right)
into asymptotic ones, when deriving (\ref{eq1}) and (\ref{eq2})
from the mother.
\par
When is it possible to make use of this equivalence, or in other words:
when can classical communication be made coherent?
The lessons learned in \cite{devetak,devetak:winter} regarding
making protocols coherent and the observations
of \cite{aram} lead us to two general rules. In what follows 
we shall work in the ``extended Hilbert space'' picture:
all quantum operations and generalized measurements are
implemented by adding ancillas (initially in pure states), 
performing unitary operations
and performing von Neumann measurements on the ancillas.
No subsystems are allowed to be discarded, so the overall 
quantum system is always in a pure state. In particular this
means that the environment $E$ is always included in our description.
Note, however, that without loss of generality a subsystem may be discarded 
after a von Neumann measurement has been performed on it; this
is because it may always be reset to a standard pure state 
via a unitary operation depending on the measurement outcome.
\par\medskip
\begin{description}
  \item[\emph{Rule I.}] If $[c \rightarrow c]$ is featured in the
    \emph{input} of a resource inequality, it 
    may be replaced by $\frac{1}{2} \left( [q \rightarrow q] - [q \, q] \right)$
    if there exists a protocol implementing the RI in
    which the classical message is almost uniformly distributed and 
    almost decoupled from the overall quantum system at the
    end of the protocol.

  \item[\emph{Rule O.}] If $[c \rightarrow c]$ is featured in the
    \emph{output} of a resource inequality with quantum inputs, it 
    may be replaced by $\frac{1}{2} \left( [q \rightarrow q] + [q \, q] \right)$
    if there exists a protocol implementing the RI in
    which the classical message is almost decoupled from the overall quantum 
    system at the end of the protocol.
    In particular, being decoupled from $E$ implies \emph{privacy}.
\end{description}
\par\medskip
In the above, a distribution $\{p_x\}$ is ``almost uniform'' when
close in trace distance to the uniform distribution.  A classical
message $x$ is ``almost independent'' of a 
quantum system $\ket{\theta_x}$  if
there exists some $\ket{\theta}$ with $\ket{\theta_x}\approx
\ket{\theta}$ for all $x$.  Throughout we write $\approx$ to denote a
trace distance of $\leq\epsilon_n$ where $\epsilon_n\rightarrow 0$ as
$n\rightarrow\infty$ for asymptotic resource inequalities (we need not
consider single-shot resource inequalities here, but the rules apply
to this case trivially with $\epsilon_n=0$).

\par
\emph{Proof of Rule I.} Whenever the resource inequality features
$[c \rightarrow c]$ in the input this means that Alice performs
a von Neumann measurement on some subsystem $A_1$, 
the outcome of which she sends to Bob,
who then performs an unitary operation depending on the received information.
Before Alice's von Neumann measurement, the joint state of  $A_1$ and 
the remaining quantum system $Q$ is
$$
\sum_x \sqrt{p_x}\ket{x}^{A_1} \ket{\phi_x}^{Q},
$$
where $p$ is an almost uniform distribution.
Upon learning the measurement outcome $x$, Bob
performs some unitary $U_x$ on $Q$, almost decoupling
it from $x$:
$$
U_x \ket{\phi_x}^{Q} = \ket{\theta_x}^{Q} \approx \ket{\theta}^Q,
$$
for some fixed state $\ket{\theta}$.  

If  Alice refrains from the measurement and
instead sends $A_1$ through a \emph{coherent} channel (\ref{map}),
the resulting state is
$$
\sum_x \sqrt{p_x} \ket{x}^{A_1} \ket{x}^{B_1} \ket{\phi_x}^{Q}.
$$
Bob now performs the \emph{controlled} unitary 
$\sum_x \proj{x}^{B_1} \otimes  U_x$, giving rise to
$$
\approx \left(\sum_x \sqrt{p_x} \ket{x}^{A_1} \ket{x}^{B_1}\right) \otimes 
\ket{\theta}^Q.
$$
Thus, in addition to the state $\ket{\theta}^Q$, 
an almost maximally entangled state has been generated. Counting
resources, $[c \rightarrow c]$ has been replaced by
$$
[q \rightarrow q \, q]  - [qq] = 
\frac{1}{2} \left( [q \rightarrow q] - [q \, q] \right).
$$
It can be shown that the uniformity condition on $p$ may
be relaxed, requiring only  
$n^{-1} \log p_x \approx {\rm const.}$ for all $x$.
\par
\emph{Proof of Rule O.} 
Now the roles of Alice and Bob are somewhat interchanged.
Alice performs a unitary operation depending on the classical
message to be sent and Bob performs a von Neumann measurement on
some subsystem $B_1$ which almost always succeeds in reproducing the message.  
Thus, before his measurement, the state of $B_1$ and 
the remaining quantum system $Q$ is
$$
\approx \ket{x}^{B_1} \ket{\phi_x}^{Q}.
$$
Based on the outcome $x$ of his measurement, Bob
performs some unitary $U_x$ on $Q$:
$$
U_x \ket{\phi_x}^{Q} = \ket{\theta_x}^Q \approx \ket{\theta}^Q,
$$
leaving the state of $Q$ almost decoupled from $x$.

Instead, Alice may perform \emph{coherent}
communication. Given a subsystem $A_1$ in the
state $\ket{x}^{A_1}$ she  encodes via 
\emph{controlled} unitary operations, yielding 
$$
\approx \ket{x}^{A_1} \ket{x}^{B_1} \ket{\phi_x}^{Q}.
$$
Bob refrains from measuring $B_1$ and instead 
performs the \emph{controlled} unitary 
$\sum_x \proj{x}^{B_1} \otimes  U_x$, giving rise to
$$
\approx \ket{x}^{A_1} \ket{x}^{B_1} \otimes \ket{\theta}^Q.
$$
By the conditions of rule O, there were no other measurements
made in the original protocol, so that the implementation of the
new coherent version is completely unitary.
Rule O follows from eq.~(\ref{eq:ccc}).
\par\medskip
The mother RI (\female) is now obtained 
from the hashing inequality (\ref{eq2}) by applying rule I. 
It can be checked that the protocol from \cite{devetak:winter}
implementing (\ref{eq2}) indeed satisfies 
the conditions of rule I. 
In this protocol the 
classical communication is used for sending a kind of ``which quantum code''
information from which the quantum information ``encoded''
is readily decoupled by ``decoding''.

The mother (\female) also follows from the noisy super-dense coding
inequality (\ref{eq3}), as implemented in \cite{H3LT}, by applying rule O.
Indeed, Eve only holds the \emph{static} purification of $\rho^{AB}$ which is
unaffected by Alice's encoding.
\par
The father RI (\male) is similarly obtained, via rule O, from (\ref{eq4}).
The main observation is that the protocol from \cite{BSST} implementing 
(\ref{eq4}) in fact outputs a \emph{private} classical channel as it is!
More precisely, in \cite{BSST} 
Alice and Bob share a maximally entangled state $\ket{\Phi_+}^{A' B'}$. 
Alice encodes her message 
$x$ via a unitary $U_x$: 
$$
x \mapsto (U_x \otimes \1) \ket{\Phi_+}^{A' B'} = 
(\1 \otimes U^*_x) \ket{\Phi_+}^{A' B'}.
$$
Applying the channel $U_\CN^{\otimes n}$ yields
$$
(\1^{BE} \otimes U^*_x) \ket{\Psi}^{BE B'},
$$
where $\ket{\Psi}^{BE B'} = U_\CN^{\otimes n}\ket{\Phi_+}^{A' B'}$.
Bob decodes $x$ inducing next to no disturbance on the quantum system
\cite{qstrong}. Finally he applies $U_x^T$ to $B'$, bringing the system $BEB'$
into the state $\ket{\Psi}$, thus decoupling it from $x$, and
justifying rule O.
\par 
Since (\ref{eq1}) is a completely new protocol, the only known implementation 
is the one we give in the paper. Therefore it can trivially be made coherent to
regenerate the mother.
The only child that cannot regenerate its parent is (5), because (QE) is
clearly an irreversible transformation.

\par
It is remarkable that comparatively simple protocols such as (\ref{eq3}) and (\ref{eq4})
can yield, via the mother and father protocols, the quantum channel capacity and
hashing inequality, respectively, which were long standing problems until very recently.
Of course, after two rounds of processing they become quite
complicated.

\paragraph{Conclusion.}
\label{sec:conclusion}
We have introduced two purely quantum coding protocols, which we showed to be
closely related to entanglement assisted coding tasks, quantum capacities
and distillability: these once long sought-after protocols descend from the
mother (\female) and father (\male) by applying teleportation or super-dense
coding. Furthermore, most of the children can be made coherent to regenerate 
their parents! What we have not shown here is that our protocols actually
give rise to information theoretically optimal resource trade-offs; 
a detailed discussion of these will be given in a forthcoming paper.

\paragraph{Acknowledgments.}
We thank C. H. Bennett, G. Burkard, J. A. Smolin and A. Ndirango 
for useful discussions. ID is partially supported by the NSA under the 
ARO grant numbers DAAG55-98-C-0041 and DAAD19-01-1-06. The latter grant
also supports AWH. AW was supported by the U.K.~Engineering and 
Physical Sciences Research Council.

\end{document}